\documentclass[prd,twocolumn,aps,amsmath,nofootinbib,superscriptaddress]
{revtex4}

\usepackage{graphicx}
\usepackage{bm}

\newcommand{\nn}{\nonumber} 
\newcommand{\bea}{\begin{eqnarray}}
\newcommand{\eea}{\end{eqnarray}}
\newcommand{\bfx}{{\bf x}}
\newcommand{\mP}{{\mathcal P}}

\begin{document}

%%%%%%%%%%%%%%%%%%%%%%%%%%%%%%%%%%%%%%%%%%
%Define Title, Author, Address, Preprint#

\preprint{CALT 68-2582} 

\title{On the generation of density perturbations at the end of inflation}

\author{Michael P.~Salem}
\affiliation{California Institute of Technology, Pasadena, CA 91125}
\email{salem@theory.caltech.edu}

%%%%%%%%%%%%%%%%%%%%%%%%%%%%%%%%%%%%%%%%%%

\begin{abstract}
Recently a mechanism was proposed whereby the primordial density perturbations 
are generated at the end of inflation.  We continue the analysis of the 
proposed model of this mechanism and calculate the maximum extent to which the 
density perturbations produced via this model can dominate over those of
the standard inflationary paradigm.  In addition, we provide a straightforward
variation of this model which allows for greater amplification of the 
density perturbations.  Finally, we show that a variation in the 
implementation of the original model results in significant non-gaussianities 
in the resulting spectrum of density perturbations.  The level of 
non-gaussianities can be made to saturate the current observational bound.
\end{abstract}


\maketitle

\section{Introduction}
\label{sec:introduction}

Measurements of the cosmic microwave background radiation \cite{cmb1}
have revealed a highly uniform background with relatively scale-free 
super-horizon perturbations on the order of a few parts in $10^5$.  A 
possible source of these perturbations is found in the quantum fluctuations
of one or more light scalar fields during an early epoch of inflation 
\cite{guth,ref}.  This is because in the (quasi-) de Sitter space 
of inflationary expansion, a quantum fluctuation in a scalar field evolves 
according to classical equations of motion after its wavelength exceeds the 
rapidly decreasing Hubble length \cite{quantfluct}.  A fluctuating scalar 
field may then be converted into an energy density perturbation via a 
variety of proposed mechanisms.

For example in the standard inflationary paradigm \cite{ref},
inflation is driven by the potential energy of a single slowly-rolling scalar 
field.  In this case the inflaton can be viewed as a unique clock 
parameterizing the evolution of the early universe.  Therefore fluctuations 
in the inflaton translate into fluctuations in the duration of inflation.  
Since the energy density of the universe redshifts more rapidly after 
inflation than during inflation, this results in energy density 
fluctuations on surfaces of constant scale factor after inflation.

More recently, it was proposed that energy density perturbations could result
from fluctuations in light scalar fields that do not contribute significantly
toward the inflationary dynamics.  For example, in the ``curvaton'' scenario 
\cite{curv} (see also \cite{other}) a light scalar field dubbed the curvaton 
receives fluctuations.  After inflation, the curvaton evolves as a massive 
fluid and therefore redshifts more slowly than the radiative products of 
reheating.  If the curvaton eventually dominates the energy density of the 
universe and then decays, density perturbations result because the duration 
of curvaton domination depends on the fluctuating curvaton.  The 
inhomogeneous reheating scenario \cite{DGZ1} also achieves density 
perturbations by varying the duration that a massive fluid dominates the 
energy density of the universe.  In this case the duration of domination is 
modulated via a decay width that depends on some fluctuating light 
scalar field.  Other ways to modulate the duration of of massive fluid's 
domination have been proposed in \cite{DGZextensions}.  These ideas have 
recently been reviewed in \cite{BTW}.

Another major variant on these proposals was recently offered in 
\cite{lyth} (note however that a similar idea was previously used in 
\cite{eoi2}).  Whereas in the standard picture the duration of inflation is 
influenced by the fluctuations in the inflaton as relevant modes leave the 
horizon, in this picture the duration of inflation is influenced by 
fluctuations in the inflaton when inflation ends.  This can occur, for 
example, if inflation ends at an inflaton value that depends upon some other 
field that receives fluctuations during inflation.  In \cite{lyth} a specific 
model was used to demonstrate that this scenario can lead to significant 
amplification of the density perturbations that result from slow-roll 
inflation.  

Here we continue the analysis of that model and calculate the 
maximum extent to which the resulting density perturbations can dominate over 
those of the standard inflationary paradigm.  We also explore the sensitivity
of this result to the tuning of model parameters.  In addition, we provide a 
straightforward variation upon this model which allows for greater 
amplification of the density perturbations.  Finally, we show that a slight 
variation in the implementation of the original model allows for significant 
non-gaussianities in the spectrum of density perturbations.  The level of 
non-gaussianities can be made to saturate the current observational bound.    

This paper is organized as follows.  In Section \ref{sec:background} we 
summarize the mechanism of \cite{lyth} and establish a convenient formalism 
and notation.  Section \ref{sec:model} describes the specific model introduced
in \cite{lyth}, and in Section \ref{sec:analysis} we analyze this model in
greater detail.  Section \ref{sec:generalization} introduces two variations
upon this model and studies their consequences.  Discussion and concluding 
remarks are provided in Section \ref{sec:conclusions}.

\section{Background}
\label{sec:background}

We parameterize metric perturbations using the curvature perturbation
on surfaces of constant scale factor, otherwise known as the Bardeen 
variable $\zeta$ \cite{bardeen}.  This corresponds to writing the interval
\bea
ds^2=-dt^2+a^2(t)e^{2\zeta(t,\bfx)}\gamma_{ij}(t,\bfx)dx^idx^j \,,
\eea
where $a$ and $\zeta$ are chosen to give $\gamma_{ij}$ unit determinant.  
In this work we ignore tensor perturbations, which allows us to write 
$\gamma_{ij}(t,\bfx)=\delta_{ij}$ since scalar field fluctuations carry no 
anisotropic stress (for a review of cosmological perturbation theory see 
\cite{FMBL}).

The proposal of \cite{lyth} is most easily studied within the so-called 
$\delta N$ formalism of \cite{dN} (see \cite{dN2} for recent extensions).  
This formalism reveals that on super-horizon scales the Bardeen variable 
$\zeta$ can be interpreted as the perturbation in the number of e-folds of 
local expansion between an ``initial'' spatially-flat hypersurface (at, say,
time $t=t_0$) and the comoving hypersurface on which $\zeta$ is evaluated 
\cite{dN},
\bea
\zeta (t,\bfx) = N(t,\bfx)-\ln\left[\frac{a(t)}{a(t_0)}\right]
\equiv\delta N(t,\bfx) \,.
\eea
On super-horizon scales, spatial gradients can be ignored and $N(t,\bfx)$
can be computed by treating large, widely separated regions as separate 
FRW universes, but with different initial field values \cite{dN}.

We limit our attention to the case where inflation is driven by a single
inflaton.  Then fluctuations in $N$ come only from the value of the inflaton
when its wavelength becomes larger than the Hubble length, denoted $\phi_k$, 
and the value of the inflaton at the end of inflation, denoted $\phi_{\rm e}$.
Thus the number of e-folds of universal expansion is 
$N(\phi_k,\phi_{\rm e})$, and
\bea
\delta N \simeq 
\frac{\partial N}{\partial \phi_k}\delta\phi_k + 
\frac{\partial N}{\partial \phi_{\rm e}}\delta\phi_{\rm e} \,.
\label{Nexp}
\eea
That the power spectrum for $\zeta$ is observed to be predominantly 
Gaussian allows us to neglect higher order terms that might appear in 
the expansion of Eq.~(\ref{Nexp}).

In the standard picture, inflation ends when the slow-roll conditions are
violated.  For single field inflation this happens at a unique value 
$\phi_{\rm e}$ and therefore the second term in Eq.~(\ref{Nexp}) is zero.  
For inflation to end at varying values of $\phi_{\rm e}$ requires 
the addition of at least one other field.  For example, with an additional 
field $\sigma$ it is possible that $\phi_{\rm e}(\sigma)$ depends upon 
$\sigma$ and inflation ends within a range of $\phi_{\rm e}$ given by
\bea
\delta\phi_{\rm e} \simeq \frac{\partial\phi_{\rm e}}{\partial\sigma}
\delta\sigma \,.
\eea

If the masses of the scalar fields are much less than the Hubble rate the 
fields acquire a constant power spectrum of fluctuations 
$\mP_{\delta\phi_k}=\mP_{\delta\sigma_k}=(H_k/2\pi)^2$, where $H_k$ is the 
Hubble rate at the time the mode $k$ exits the horizon \cite{fluctrev}.  
Thus the power spectrum for $\zeta$ is 
\bea
\mP_\zeta \simeq \frac{H_k^2}{4\pi^2} \left[ 
\left( \frac{\partial N}{\partial \phi_k} \right)^2 \!\! + 
\left( \frac{\partial N}{\partial \phi_{\rm e}} 
\frac{\partial \phi_{\rm e}}{\partial \sigma_k} \right)^2 \right] \,.
\label{PZ}
\eea
The density perturbations generated at the end of inflation will  
dominate over those produced from the standard picture when
\bea
\frac{\partial N}{\partial \phi_{\rm e}} 
\frac{\partial \phi_{\rm e}}{\partial \sigma} >
\frac{\partial N}{\partial \phi_k} \,.
\label{domcond}
\eea
Note that 
$\partial N/\partial \phi_k \neq \partial N /\partial \phi_{\rm e}$
since the former probes the dependence of $N$ on $\phi$ deep in the 
inflationary epoch while the latter probes the dependence of $N$ on 
$\phi$ near the end of inflation.

\section{The Specific Model}
\label{sec:model}

The above ideas were demonstrated in \cite{lyth} using a specific model 
described by the potential
\bea
V(\phi,\chi,\sigma) \!&=&\! \frac{1}{4g}\left( m_\chi^2-g\chi^2 \right)^2
+\frac{1}{2}m_\phi^2\phi^2+\frac{1}{2}\lambda_\phi\phi^2\chi^2 \nn\\
\!& &\! + \, \frac{1}{2}\lambda_\sigma\sigma^2\chi^2 +V_\sigma (\sigma) \,.
\label{V}
\eea
As was noted in \cite{lyth}, this is the original hybrid inflation model 
\cite{hybrid} but with interactions involving an additional light scalar 
field $\sigma$ added with the last two terms.  

Hybrid inflation assumes the initial conditions of chaotic inflation 
\cite{chaotic,hybrid}.  However, during the early stages of hybrid inflation 
the field $\chi$ is pushed to zero much faster than $\phi$ (and likewise faster
than $\sigma$ in the above model).  Then $\phi$ becomes a slowly-rolling 
inflaton with inflation assisted by the vacuum energy $m_\chi^4/4g$.  As in
\cite{lyth} we assume that the self interaction of $\sigma$ represented by
the term $V_\sigma$ does not contribute significantly toward the inflationary 
dynamics.

The field $\chi$ is pinned to the origin until the curvature in its 
potential, $\partial^2V/\partial\chi^2$, becomes negative.  In \cite{lyth}
a scenario is described in which inflation ends abruptly as  $\chi$ rolls 
away from the origin.  This happens when 
\bea
\frac{\partial^2V}{\partial\chi^2}=
\lambda_\phi\phi_{\rm e}^2 +\lambda_\sigma\sigma_{\rm e}^2-m_\chi^2 =0 \,.
\label{eoic}
\eea
Therefore if $\sigma$ receives fluctuations $\delta\sigma$, inflation ends
at field values $\phi_{\rm e}$ that vary according to 
\bea
\delta\phi_{\rm e} \simeq 
-\frac{\lambda_\sigma\sigma_{\rm e}}{\lambda_\phi\phi_{\rm e}} \delta\sigma \,.
\eea
Referring back to Eq.~(\ref{PZ}), we see the density perturbations resulting
from fluctuations $\delta\phi_{\rm e}$ dominate over those resulting from 
$\delta\phi_k$ when
\bea
R\equiv \frac{\lambda_\sigma^2\sigma_{\rm e}^2}{\lambda_\phi^2\phi_{\rm e}^2}
\left( \frac{\partial N/\partial \phi_{\rm e}}{\partial N/\partial\phi_k}
\right)^2 = 
\frac{\lambda_\sigma^2\sigma_{\rm e}^2}{\lambda_\phi^2\phi_{\rm e}^2}
\frac{\epsilon_k}{\epsilon_{\rm e}} > 1 \,.
\label{R}
\eea
In the second equation we have used that the first slow-roll parameter can 
be written
\bea
\epsilon\equiv \frac{m_{\rm pl}^2}{2}
\left( \frac{1}{V}\frac{\partial V}{\partial\phi} \right)^2
= \frac{1}{2m_{\rm pl}^2}
\left( \frac{\partial N}{\partial\phi} \right)^{-2} \,.
\eea

An important consequence of this mechanism is that the spectral tilt is 
given by the tilt in the spectrum of fluctuations $\delta\sigma$, as opposed
to $\delta\phi$.  This gives a tilt which is independent of the second 
slow-roll parameter $\eta$ \cite{lyth,fluctrev}.  Therefore $\eta$ is not 
directly constrained by observation.  In addition, since 
$\dot{\epsilon}\simeq 2H\epsilon\,(2\epsilon-\eta)$ it appears as if 
$\epsilon$ may decrease significantly during the course of inflation if 
$\eta\gg \epsilon$.  Thus it was suggested in \cite{lyth} that the condition 
of Eq.~(\ref{R}) is easily satisfied.

However, one might note that $R$ is proportional to $\epsilon_k$ which 
according to this mechanism is constrained by observational bounds on the 
spectral tilt.  These give $\epsilon_k\lesssim 0.02$ \cite{cmb1} in the 
absence of cancellations \cite{lyth}.  In addition, we do not expect 
$\epsilon_{\rm e}$ and $\phi_{\rm e}$ to be independent of each other.  Thus 
it is worthwhile to investigate Eq.~(\ref{R}) in greater detail in order to 
determine precisely what limits the extent to which the density perturbations 
of this model can dominate over those produced via the standard inflationary 
paradigm.

\section{A More Detailed Analysis}
\label{sec:analysis}

We now consider the model of Eq.~(\ref{V}) in greater detail in order to 
clarify the constraints on $R$ as given by Eq.~(\ref{R}).  Consistent
with the analysis of \cite{lyth} we ignore the contribution of $V_\sigma$ 
toward the vacuum energy.  As described above, the $\chi$ field rolls away 
from the origin when $\phi$ takes the value
\bea
\phi_{\rm e} = 
\sqrt{\frac{m_\chi^2-\lambda_\sigma\sigma_{\rm e}^2}{\lambda_\phi}}
\approx \frac{m_\chi}{\sqrt{\lambda_\phi}} \,.
\label{pe}
\eea
To explain the second approximation, first note that we require 
$\lambda_\sigma\sigma_{\rm e}^2<m_\chi^2$ in order to avoid $\chi$ remaining 
trapped at the origin after $\phi$ reaches zero.  Meanwhile, the precise 
value of $\sigma_{\rm e}$ is a stochastic variable constrained by conditions
independent of $\lambda_\sigma$ and $m_\chi$.  Thus to ensure that the 
desired dynamics are typical of this model requires we set 
$m_\chi^2/\lambda_\sigma \gg \sigma_{\rm e}^2$ for typical values of 
$\sigma_{\rm e}$.

For inflation to last until $\phi=\phi_{\rm e}$ but end abruptly when 
$\chi$ rolls away from the origin requires that the term $m_\chi^4/4g$ 
dominate the energy density at the end of inflation and that $m_\chi^2$ be 
much greater than the Hubble rate at the end of inflation.\footnote{The
other possibility is to set $\phi_{\rm e}=m_\chi/\sqrt{\lambda_\phi}$
to be greater than the value of $\phi$ for which inflation ends when
the potential of $\phi$ dominates.  Since in this case inflation ends when
$\phi$ is of the order of the Planck mass \cite{chaotic}, this requires a 
Planck scale $m_\chi$ unless $\lambda_\phi\gg 1$.  Nevertheless, an 
analysis similar to that which follows gives
\bea
R\simeq \frac{\phi_k^2}{\sigma_{\rm e}^2}
\bigg( \frac{\lambda_\sigma\sigma_{\rm e}^2}{m_\chi^2}\bigg)^2 \,,
\eea
where $\phi_k$ is $\phi$ evaluated 60 e-folds before the end of inflation.  
It will be seen that this is no better than the result obtained in
the scenario analyzed in greater detail in what follows.}  
These conditions respectively give the constraints
\bea
\frac{2g}{\lambda_\phi}\frac{m_\phi^2}{m_\chi^2} \ll 1 \,, \quad
\frac{1}{12g}\frac{m_\chi^2}{m_{\rm pl}^2} \ll 1 \,,
\label{constr1}
\eea
where $m_{\rm pl}$ denotes the reduced Planck mass.
The first constraint of Eqs.~(\ref{constr1}) allows us to write
\bea
\frac{\partial N}{\partial\phi_{\rm e}} = \frac{V}{m_{\rm pl}^2} 
\left( \frac{\partial V}{\partial\phi_{\rm e}} \right)^{-1} \!\!\! \simeq
\frac{m_\chi^4}{4g\phi_{\rm e} m_\phi^2m_{\rm pl}^2} \approx
\frac{\sqrt{\lambda_\phi}m_\chi^3}{4gm_\phi^2m_{\rm pl}^2} \,. \,\, 
\eea 
Meanwhile 
\bea
\frac{\partial N}{\partial\phi_k} =
\frac{1}{\phi_k m_\phi^2m_{\rm pl}^2} \left( 
\frac{m_\chi^4}{4g}+\frac{1}{2}m_\phi^2\phi_k^2 \right) \,.
\eea
Plugging these values into Eq.~(\ref{R}) gives
\bea
R \approx \frac{\lambda_\sigma}{4g^2}
\left( \frac{\lambda_\sigma\sigma_{\rm e}^2}{m_\chi^2} \right) 
m_\chi^6\phi_k^2 \left( \frac{m_\chi^4}{2g}+m_\phi^2\phi_k^2 \right)^{-2} .
\label{Rpk}
\eea
The quantity $\phi_k$ is the value of the inflaton when the mode $k$ 
exits the particle horizon.  For cosmological scales of current interest 
this happens about 60 e-folds of inflation prior to the end of inflation.

In order to find the maximum value for $R$, we find the value of $\phi_k$ 
that maximizes $R$ and set the parameters $g$, $m_\chi$, and $m_\phi$ 
such that the mode $k$ exits the horizon 60 e-folds prior to inflation.  
This is equivalent to finding the balance between the vacuum energy 
$m_\chi^4/4g$ and the potential energy $m_\phi^2\phi^2$ that maximizes
$R$.  The result is   
\bea
\phi_k = \frac{1}{\sqrt{2g}}\frac{m_\chi^2}{m_\phi} \,,
\label{pk}
\eea 
with the constraint that the mode $k$ leaves the horizon at $N_k\approx 60$, 
with 
\bea
N_k \!&=&\! \frac{1}{4m_\phi^2m_{\rm pl}^2}\left[ 
\frac{m_\chi^4}{g}\ln\left(\frac{\sqrt{\lambda_\phi}\phi_k}{m_\chi}\right)
+m_\phi^2\phi_k^2-\frac{m_\phi^2m_\chi^2}{\lambda_\phi} \right] \nn\\
\!&\approx&\! \frac{m_\chi^4}{8gm_\phi^2m_{\rm pl}^2}\ln\left( 
\frac{\lambda_\phi}{2g}\frac{m_\chi^2}{m_\phi^2}\right) \,.
\label{Nk}
\eea
In the last expression we have used the constraints of 
Eqs.~(\ref{constr1}) to identify the most significant term.  Putting this 
all together we find the maximum value of $R$ to be 
\bea
R\approx N_k\,\bigg(\frac{\lambda_\sigma\sigma_{\rm e}^2}{m_\chi^2}\bigg)^2
\frac{m_{\rm pl}^2}{\sigma_{\rm e}^2}
\left[ \ln \bigg( \frac{\lambda_\phi}{2g}\frac{m_\chi^2}{m_\phi^2} 
\bigg) \right]^{-1} .
\label{Rdefinate}
\eea   

Before proceeding to study Eq.~(\ref{Rdefinate}) we should check that 
producing the correct power spectrum normalization does not introduce 
any constraints that conflict with our present assumptions.  Applying 
Eq.~(\ref{PZ}) to the scenario considered above we find that when 
density perturbations generated at the end of inflation dominate we have
\bea
\mP_\zeta \approx \frac{2}{\pi^2} N_k R \,
\bigg( \frac{1}{12g}\frac{m_\chi^2}{m_{\rm pl}^2} \bigg)
\frac{m_\chi^2}{m_{\rm pl}^2}
\left[ \ln \bigg( \frac{\lambda_\phi}{2g}\frac{m_\chi^2}{m_\phi^2} 
\bigg) \right]^{-1} . \,
\eea  
To match observation we require $\mP_\zeta$ to be very small, 
$\mP_\zeta \approx (5\times 10^{-5})^2$ \cite{cmb1}.  According to the second 
of Eqs.~(\ref{constr1}) the first term in parentheses is already constrained 
to be much less than order unity and in fact can be set as small as necessary 
to match observation.  In addition we expect $m_\chi^2/m_{\rm pl}^2$ to be 
very small.  Thus to set $\mP_\zeta$ to match observation does not introduce 
any constraints in conflict with those of the above analysis.  

We have written $R$ in the form of Eq.~(\ref{Rdefinate}) in order to 
emphasize the maximum extent to which the perturbations produced at the 
end of inflation may dominate over those produced as cosmological scales
exit the horizon.  The first term is weakly constrained by the energy scale 
of inflation and is here taken to be $N_k\approx 60$.  The first term in 
parentheses is constrained to be significantly less than unity as explained 
in the discussion below Eq.~(\ref{pe}).  Finally, the argument of the 
logarithm must be much greater than unity in order to satisfy the first 
constraint of Eqs.~(\ref{constr1}).  Therefore the last term is less than 
unity.  Thus for appropriately tuned parameters we might expect the product
of these three factors to be a couple orders of magnitude below unity.

However, Eq.~(\ref{Rdefinate}) also contains a factor 
$m_{\rm pl}^2/\sigma_{\rm e}^2$.  In (quasi-) de Sitter space a scalar field
such as $\sigma$ evolves both according to its classical equation of motion
and due to quantum fluctuations as modes leave the particle horizon.  The 
net effect of this evolution is that the correlation function 
$\langle\sigma^2\rangle$ migrates toward a fixed value depending upon 
$V_\sigma$ and the Hubble rate $H$ \cite{css}.  For example, if inflation 
lasts long enough and if $V_\sigma=\frac{1}{2}m_\sigma^2\sigma^2$ then 
$\langle\sigma^2\rangle\sim H^4/m_\sigma^2$ \cite{css}.  Taking a typical
value of $\sigma_{\rm e}$ to be 
$\sigma_{\rm e}\sim \sqrt{\langle\sigma^2\rangle}$ gives 
\bea
\frac{m_{\rm pl}^2}{\sigma_{\rm e}^2} \sim 
\frac{m_\sigma^2}{H^2}\frac{m_{\rm pl}^2}{H^2} \,.
\label{sigmaterm}
\eea
The dynamics described above require $m_\sigma^2\ll m_\phi^2\ll H^2$, so the 
first term in Eq.~(\ref{sigmaterm}) must be at least a few orders of 
magnitude below unity.  However, the second term can be very large.  Current 
observation gives $m_{\rm pl}^2/H^2 \gtrsim 10^8$ \cite{cmb2}, which 
is more than sufficient to compensate for all the small factors in $R$ if 
parameters are tuned appropriately.  Reducing the scale of inflation allows 
for greater values of $R$.  Of course, our Hubble volume could also be a 
region of atypically small $\sigma_{\rm e}$.  Finally, we may reduce 
$\sigma_{\rm e}$ to an arbitrarily small scale by considering $\sigma$ to be 
a pseudo-Nambu-Goldstone boson (see for example \cite{curv,DLNR}).\footnote{
An interesting scenario involves a pseudo-Nambu-Goldstone boson that ranges 
over a scale $\sim m_\chi$.  Then $\lambda_\sigma$ may be a coupling of order 
unity and we obtain
\bea
R\approx N_k\frac{m_{\rm pl}^2}{m_\chi^2} \,,
\eea
where we have dropped the logarithm and other factors of order unity.  
Clearly $R$ is much greater than unity in this case.}
Thus in a number of circumstances we expect the level of density perturbations 
generated at the end of inflation to be significantly larger than those 
produced when cosmological scales exit the horizon. 

In Eq.~(\ref{Rdefinate}) the parameters $g$, $m_\phi$, and $m_\chi$ are tuned 
such that scales of cosmological interest leave the horizon when $\phi_k$ is 
given by Eq.~(\ref{pk}).  To study the result of relaxing this tuning 
requires to invert Eq.~(\ref{Nk}) to obtain $\phi_k(N_k)$ and insert the 
result into Eq.~(\ref{Rpk}) to obtain $R\,(N_k)$.  This allows for the 
remaining parameters in $R$ to be freely varied while $R$ retains its original
meaning; that is, that $R$ compares density perturbations produced at the 
end of inflation to those produced when cosmological scales exit the 
particle horizon.  

Note that the first two terms in the brackets of Eq.~(\ref{Nk}) always 
dominate over the third term and that they are comparable to each other when 
$\phi_k$ is given by Eq.~(\ref{pk}).  Remember that $\phi_k$ is defined as 
the value of $\phi$ $N_k\approx 60$ e-folds before the end of inflation.  
Decreasing $m_\phi$ slows the evolution of $\phi$ which therefore decreases 
$\phi_k$.  In this case the first term in brackets becomes more important and 
inverting Eq.~(\ref{Nk}) gives
\bea
\phi_k^2\approx \phi_{\rm e}^2 \exp \bigg( 8N_k
\frac{gm_\phi^2m_{\rm pl}^2}{m_\chi^4}\bigg) \,.
\eea
For $\phi_k$ less than in Eq.~(\ref{pk}) the important functional dependence 
of $R$ is $R\,(\phi_k)\propto \phi_k^2$.  Thus we see that $R$ decreases 
exponentially when the ratio $gm_\phi^2m_{\rm pl}^2/m_\chi^2$ is decreased 
from its optimal value.  

On the other hand, increasing $m_\phi$ quickens the evolution of $\phi$ and 
therefore increases $\phi_k$.  In this case the second term in the brackets 
of Eq.~(\ref{Nk}) becomes more important.  Inverting $N_k$ in this case gives 
$\phi_k^2\approx N_k m_{\rm pl}^2$ which is relatively independent of the 
model parameters.  Therefore the magnitude of $R$ changes predominantly 
through its dependence upon $m_\phi^2\phi_k^2$ in the denominator of 
Eq.~(\ref{Rpk}).  Thus we see that significantly increasing the ratio 
$g m_\phi^2m_{\rm pl}^2/m_\chi^4$ from its optimal value results in a roughly 
proportional decrease in the in size of $R$.

\section{Generalizing the Model}
\label{sec:generalization}

\subsection{Varying the Potential for $\phi$}

According to Eq.~(\ref{domcond}), the generation of density perturbations
at the end of inflation is most effective when the slow-roll parameter near 
the end of inflation is much less than when cosmological scales of interest 
exit the particle horizon.  Thus to enhance the resulting perturbations we 
desire a potential for $\phi$ that decreases more steeply for $\phi$ deep in 
the inflationary epoch and decreases more gently for $\phi$ near the end of 
inflation.  This can be accomplished by replacing the 
$\frac{1}{2}m_\phi^2\phi^2$ term in Eq.~(\ref{V}) with 
$\frac{1}{4}\lambda\phi^4$.    

We analyze this scenario in exact analogy to the analysis in Section 
\ref{sec:analysis}.  Again we assume that the vacuum energy at the end of 
inflation is dominated by the $m_\chi^4/4g$ term.  The calculations proceed
just as in Section \ref{sec:analysis}, and in the end we find
\bea
R \simeq 108N_k^3 
\bigg( \frac{\lambda_\sigma\sigma_{\rm e}^2}{m_\chi^2} \bigg)^2
\bigg(\frac{m_{\rm pl}^2}{\sigma_{\rm e}^2}\bigg)
\bigg( \frac{\lambda_\phi m_{\rm pl}^2}{m_\chi^2} \bigg)^2 \,.
\eea   
As described in Section \ref{sec:analysis}, the term in the first set of 
parentheses is expected to be significantly less than unity.  However the 
last term is expected to be greater than unity in order that $\chi$ becomes 
pinned to the origin in the early stages of inflation.  In addition, the 
numerical pre-factor $108N_k^3\sim 10^7$ for $N_k\approx 60$.  The 
discussion about the factor $m_{\rm pl}^2/\sigma_{\rm e}^2$ near the end of 
Section \ref{sec:analysis} applies here.  Thus we see the maximum level of
density perturbations produced at the end of inflation can be greatly 
amplified by simply introducing a $\phi^4$ potential.

\subsection{Relaxing the Constraint on $\lambda_\sigma$}
\label{ssec:relax}
\footnote{This subsection fails to account for the lack of correlation 
between $\sigma$ and $\phi$ fluctuations when computing the 
non-gaussianity parameter $f_{\rm NL}$.  The issue is discussed in the 
erratum.}

In Section \ref{sec:model} we followed the assumption presented in 
\cite{lyth} that near the end of inflation 
$\lambda_\sigma\sigma^2<m_\chi^2$ so that inflation ends before 
$\phi$ reaches zero.  It is interesting to explore the consequences of 
lifting this assumption.  If we do not demand that 
$\lambda_\sigma\sigma^2<m_\chi^2$, then for large enough $\lambda_\sigma$ 
or $\sigma$ the $\chi$ field is still pinned to the origin when 
$\phi$ reaches zero.  We then expect inflation to continue with the $\sigma$ 
field rolling down its potential until the condition of Eq.~(\ref{eoic}) is 
met.  At this point $\chi$ rolls away from the origin and abruptly initiates 
the end of inflation.  

Nevertheless, if $m_\phi \ll H$ during inflation, then $\phi$ will retain a 
power spectrum of fluctuations given by $\mP_{\delta\phi_k}=(H_k/2\pi)^2$.
Thus the fields $\sigma$ and $\phi$ have essentially changed places, with
$\sigma$ playing the part of the inflaton and $\phi$ the fluctuating field.
However, in this scenario $\phi$ has no homogeneous component and 
\bea
\delta\sigma_{\rm e} = -\frac{\lambda_\phi}{\lambda_\sigma}
\frac{\delta\phi_k^2}{\sigma_{\rm e}} \,.
\eea
The density perturbations produced at the end of inflation are now 
entirely non-gaussian!  

It should be noted here that the original model described in Section 
\ref{sec:model} also results in non-gaussianities.  As described in 
\cite{lyth}, in that model the non-gaussianities result from higher order 
terms in the expansion of $\delta N$ in Eq.~(\ref{Nexp}).  The situation we 
consider here is different in that the non-gaussianities arise at leading 
order in the expansion of $\delta N$.  

The level of non-gaussianity can be parameterized with a term $f_{\rm NL}$ 
defined according to the equation
\bea
\zeta = \zeta_{\rm g} - \frac{3}{5}f_{\rm NL}\zeta_{\rm g}^2 \,,
\label{fnldef}
\eea
where $\zeta_{\rm g}$ symbolizes a variable with a gaussian spectrum 
\cite{nonguass}.  In this case $\zeta_{\rm g}$ is the curvature perturbation 
produced via the standard inflationary paradigm,  
\bea
\zeta_{\rm g} = \frac{\partial N}{\partial\sigma_k}\delta\sigma_k \,.
\label{zg}
\eea
Here we have assumed for simplicity that scales of cosmological interest 
exit the horizon after $\phi$ has rolled to zero.  Thus we treat the 
relevant stages of inflation as driven by the $\sigma$ field, assisted 
by the vacuum energy $m_\chi^4/4g$, with $\phi$ being a heavier field
fluctuating about the origin.  We neglect the non-gaussian component 
of $\zeta$ produced when relevant scales exit the horizon as this is in 
general relatively small \cite{maldacena}.  For comparison to observation 
it does not matter whether the non-gaussian fluctuations are sourced by the 
same field as the gaussian fluctuations \cite{ngcmb}.  Therefore we can 
combine Eq.~(\ref{Nexp}), Eq.~(\ref{fnldef}) and Eq.~(\ref{zg}) to obtain
\bea
f_{\rm NL} \!&=&\! - \frac{5}{3}
\left( \frac{\partial N}{\partial\sigma_{\rm e}}\delta\sigma_{\rm e}\right)
\left( \frac{\partial N}{\partial\sigma_k}\delta\sigma_k \right)^{-2} \nn\\
\!&=&\! \frac{5}{3} \frac{\lambda_\phi}{\lambda_\sigma}
\frac{1}{\sigma_{\rm e}}\frac{\partial N}{\partial\sigma_{\rm e}}
\left( \frac{\partial N}{\partial\sigma_k} \right)^{-2} \,.
\label{f2}
\eea
If we take $V_\sigma=\frac{1}{2}m_\sigma^2\sigma^2$, it is straightforward 
to calculate the largest possible $f_{\rm NL}$ by translating the arguments 
of Section \ref{sec:analysis}.  In particular, we note that 
$\sigma_{\rm e}=m_\chi/\sqrt{\lambda_\sigma}$ and 
that 
\bea
\frac{\partial N}{\partial\sigma_{\rm e}} \simeq \frac{1}{4g}
\frac{m_\chi^4}{\sigma_{\rm e}m_\sigma^2m_{\rm pl}^2} = 
\frac{\sqrt{\lambda_\sigma}}{4g}\frac{m_\chi^3}{m_\sigma^2m_{\rm pl}^2} \,.
\eea
Likewise, minimizing the factor $(\partial N/\partial\sigma_k)^{-2}$ gives 
$\sigma_k=m_\chi^2/\sqrt{2g}m_\sigma$ and 
\bea
\frac{\partial N}{\partial\sigma_k} = \frac{1}{\sqrt{2g}}
\frac{m_\chi^2}{m_\sigma m_{\rm pl}^2} \,.
\label{dNds}
\eea
Finally, putting all this together we find 
\bea
f_{\rm NL} \simeq \frac{5}{6}\frac{\lambda_\phi m_{\rm pl}^2}{m_\chi^2} \,.
\label{fnl}
\eea
It must be emphasized this is an upper limit; smaller $f_{\rm NL}$ are 
easily achieved by choosing parameters that do not minimize 
$(\partial N/\partial\sigma_k)^{-2}$.  Note also that in this implementation
of the model the only constraint on $\lambda_\phi$ is that 
$\lambda_\phi \delta\phi^2=\lambda_\phi (H_k/\sqrt{2})^2 \ll m_\chi^2$.  This
constraint assures that $\chi$ always rolls away from the origin before
$\sigma$ reaches zero, and gives
\bea
\lambda_\phi\bigg( \frac{1}{12g}\frac{m_\chi^2}{m_{\rm pl}^2} \bigg) \ll 1 \,.
\eea
Referring to second constraint of Eqs.~(\ref{constr1}), we see the term
in parenthesis is already constrained to be much less than unity.  Therefore 
this mechanism permits a non-gaussian component to the density perturbations 
all the way through the observational limit of 
$f_{\rm NL}\lesssim 135$ \cite{ngcmb}.    

For the above calculation to be appropriate requires that the 
$\zeta_{\rm g}$ of Eq.~(\ref{zg}) actually be the dominant contribution to
the curvature perturbation.  For this to be the case its power spectrum 
must have a magnitude $\mP_{\zeta,{\rm g}}\approx (5\times 10^{-5})^2$ in 
order to match observation \cite{cmb1}.  From Eq.~(\ref{PZ}) and 
Eq.~(\ref{dNds}) we have
\bea
\mP_{\zeta,{\rm g}} = 
\left( \frac{\partial N}{\partial\sigma_k}\right)^2\frac{H_k^2}{4\pi^2}
= \frac{1}{8\pi^2\epsilon_k} \frac{H_k^2}{m_{\rm pl}^2} \,,
\eea
where we have used that in this scenario the first slow-roll parameter as 
the mode $k$ exits the horizon is 
\bea
\epsilon_k = \frac{1}{2m_{\rm pl}^2} \left( 
\frac{\partial N}{\partial\sigma_k} \right)^{-2} 
\!\!= \frac{g\,m_\sigma^2 m_{\rm pl}^2}{m_\chi^4} \,.
\label{e}
\eea
We have written the power spectrum in this way in order to employ the
observational constraints on the spectral tilt and on the level of 
gravity waves.  

Since the primary, gaussian density perturbations are now sourced during
inflation, the spectral tilt is 
$n-1=2\eta-6\epsilon_k \approx 2\epsilon_k \lesssim 0.04$ \cite{cmb1}, 
where we have used that the second slow-roll parameter in the scenario we 
are considering is given by
\bea
\eta \equiv \frac{m_{\rm pl}^2}{V}\frac{\partial^2 V}{\partial\sigma^2}
\approx \frac{4g\,m_\sigma^2m_{\rm pl}^2}{m_\chi^4} \approx 4\epsilon_k \,.
\label{eta}
\eea
Meanwhile, it is observed that $H_k^2/m_{\rm pl}^2\lesssim 10^{-8}$
\cite{cmb2}.  In this model we can decrease the spectral tilt independently
of $H_k^2/m_{\rm pl}^2$ by simply decreasing $m_\sigma$.  Therefore we 
simply require to set $H_k^2/m_{\rm pl}^2\lesssim 10^{-9}$ to match 
observation.  Here
\bea
\frac{H_k^2}{m_{\rm pl}^2} = \frac{1}{6g}\frac{m_\chi^4}{m_{\rm pl}^4}
= \frac{5}{3}\lambda_\phi f_{\rm NL}^{-1} 
\bigg( \frac{1}{12g} \frac{m_\chi^2}{m_{\rm pl}^2} \bigg) \,,
\label{Hlimit}
\eea        
where we have expressed the result in terms of $f_{\rm NL}$ to clarify
which are the remaining free parameters.  According to the second 
constraint in Eq.~(\ref{constr1}), the term in parentheses is already 
constrained to be much less than unity.  In fact, it can be set as small as 
necessary to satisfy observation.  Moreover, since the three terms in 
Eq.~(\ref{Hlimit}) depend upon three independent parameters 
($\lambda_\phi$, $g$, and $m_\chi$), we have considerable freedom in
exactly how we satisfy the observational bound.  Thus we conclude that this 
model allows for significant non-gaussianities for a range of model 
parameters.

\section{Discussion and Conclusions}
\label{sec:conclusions}

In this work we continue the analysis of the model proposed in \cite{lyth} 
to generate density perturbations at the end of inflation.  We confirm that 
these density perturbations can easily dominate over those produced via the 
standard inflationary paradigm, and explore the sensitivity of this 
result to the tuning of model parameters.  In addition, we provide a 
straightforward variation of this model which allows for even greater 
amplification of the density perturbations.

It is worthwhile to consider how general is this analysis.  According to 
Eq.~(\ref{domcond}) the production of density perturbations at the end of 
inflation is most effective when the slow-roll parameter near the end of 
inflation is much less than that when cosmological scales of interest exit 
the particle horizon.  Since inflation can only end with the slow-roll 
parameter rising to unity, this suggests the mechanism is most effective 
only when the inflationary potential contains large derivatives near the 
point where it dips toward its minimum.  These large derivatives are most 
naturally accomplished by inserting a second field direction for the vacuum 
energy to fall to zero and initiate a reheating phase.  This is precisely 
the scenario implemented in hybrid inflation and generalized in the models 
considered here.

We also study a variation in the implementation of the model proposed in 
\cite{lyth} that results in modified inflationary dynamics.  We show that 
this case results in a spectrum of density perturbations with significant 
non-gaussianities for a range of model parameters.  In particular it is 
shown that these non-gaussianities are capable of saturating the current 
observational bound.

\section{Erratum}

In Eq.~(\ref{fnldef}) we define the non-gaussianity parameter 
$f_{\rm NL}$ and in Eq.~(\ref{f2}) we relate it to model parameters.
This definition and the subsequent calculation is appropriate when the 
gaussian and non-gaussian contributions to $\zeta$ come from fluctuations 
in the same field, or when they are correlated as such, but when they
are uncorrelated, a different approach is more appropriate~\cite{L2}.  
What is defined as $f_{\rm NL}$ above corresponds to a quantity
\bea
f_{\rm NL} = -\frac{5}{3}r_\zeta \mP_\zeta^{-1/2} \,,
\label{ef}
\eea
where $r_\zeta$ is the non-gaussian fraction of Ref.~\cite{L2}.  
As explained in Ref.~\cite{L2}, the non-gaussianity parameter that is
relevant to observation is given by
\bea
\tilde{f}_{\rm NL} \sim -r_\zeta^3\mP_\zeta^{-1/2} \,,
\eea   
where some factors of order unity and a logarithmic dependence on the 
scale of observation $k$ have been suppressed.  Combining with 
Eq.~(\ref{ef}) gives
\bea
\tilde{f}_{\rm NL} \sim f_{\rm NL}^3 \mP_\zeta \,.
\eea
This result is suppressed by a factor 
$\mP_\zeta\approx (5\times 10^{-5})^2$, but includes the quantity 
$f_{\rm NL}$ of Eq.~(\ref{fnl}) cubed.  Therefore, whereas it is argued 
in Section~\ref{ssec:relax} that the right-hand side of Eq.~(\ref{fnl})
need only approach ${\cal O}(100)$ to reach the observational limit, we 
here find it must approach ${\cal O}(10^4)$ to reach this limit.  Still, 
this appears to be easy to achieve in this model.  Note also that these 
results assume the vacuum expectation value of the light field $\phi$ 
is precisely zero (no homogeneous component), as is the correlation 
between $\phi$ and $\sigma$.  Violation of these assumptions (due for 
instance to averaging over just one Hubble volume) may lead to increased 
levels of non-gaussianity for the same parameter values.

The author is grateful to Christian Byrnes for explaining these issues
to him.

\begin{acknowledgments}
MS thanks Michael Graesser and Mark Wise for helpful discussions, Michael 
Graesser for useful comments on the manuscript, David Lyth for correcting 
an assumption made in the first report of this work and for other helpful 
comments, Misao Sasaki for pointing out some additional misstatements, 
and Christian Byrnes for pointing out the problem with 
Section~\ref{ssec:relax}.  This work was supported by the DoE under 
contract DE-FG03-92ER40701.    
\end{acknowledgments}

\end{document}